\documentclass[preprint,samsmath,amssymb,floatfix,superscriptaddress,nofootinbib,aps]{revtex4}
\usepackage{graphicx}
\begin{document}
\title{
Ground state clusters for short-range attractive \\
and long-range repulsive potentials
}
\def\roma{\affiliation{
Dipartimento di Fisica and INFM Udr and SOFT: Complex Dynamics in 
Structured Systems, Universit\`a  di Roma ``La Sapienza'', 
P.le A. Moro 2, I-00185, Roma, 
Italy}}
\def\smc{\affiliation{
Dipartimento di Fisica and INFM Udr and SMC: Center for Statistical 
Mechanics and Complexity, Universit\`a  di Roma ``La Sapienza'', 
P.le A. Moro 2, I-00185, Roma, Italy}}
\def\esrf{\affiliation{
European Synchrotron Radiation Facility, B.P. {\em 220}, 
F-{\em 38043} Grenoble,
Cedex France
}}
\author{S.~Mossa}\roma\esrf
\author{F.~Sciortino}\roma
\author{P.~Tartaglia}\smc
\author{E.~Zaccarelli}\roma
\date{\today}
\begin{abstract}
We report calculations of the ground state energies and geometries for
clusters of different sizes (up to 80 particles), where individual
particles interact simultaneously via a short-ranged attractive 
--- modeled with a generalization of the Lennard-Jones potential --- 
and a long-ranged repulsive Yukawa potential. We show that, for specific
choices of the parameters of the repulsive potential, the ground state
energy per particle has a minimum at a finite cluster size. For these
values of the parameters in the thermodynamic limit, at low
temperatures and small packing fractions --- where clustering is
favored and cluster-cluster interactions can be neglected --- 
thermodynamically stable cluster phases can be formed. The analysis
of the ground state geometries shows that the spherical shape is
marginally stable. In the majority of the studied cases, we find that,
above a certain size, ground state clusters preferentially grow 
almost in one dimension.
\end{abstract}
\maketitle
\section{Introduction}
\label{intro:sec}
Understanding the formation of self-assembled structures at nano and
mesoscopic level is one of the central issues in condensed matter
studies~\cite{review-self-assembly}, and particularly in biological and
soft matter fields. More recently, significant efforts have been made
in the direction of finding connections between features of the
inter-particle potential and the resulting stable supra-particles
patterns. In this respect, an interesting class of systems in which
particles self assemble into large aggregates is generated by systems 
of particles interacting simultaneously via an attractive potential 
(which can be of van der Walls or depletion origin) and a repulsive 
interaction (usually of screened electrostatic origin). 
In density and temperature conditions where the attractive part of the 
potential would generate macroscopic phase separations, the presence of 
a repulsive part contrasts the phase separation process, leaving the system, 
in some cases, in a microphase separated condition. 
The competition between attraction and repulsion generates the formation of 
equilibrium phases where stable clusters of particles are formed (sometime 
called cluster or micellar phases) or even of more complex structures 
(such as lamellar or columnar phases). 
Experimental evidence for such thermodynamically stable cluster phases 
have been recently presented for colloidal systems and for solutions of proteins at 
low ionic strength~\cite{chenmessina,schurtenberger,prlweitz,daveweitz,poon-exp-gel}. 
Cluster phases have also been observed in aqueous solutions of silver 
iodide~\cite{kegel2}. Theoretically, the formation of phases with different 
morphology has been addressed within lattice models and mean field calculations 
for unscreened electrostatic 
interactions~\cite{groussonpre,lowprl,schmalianprl,chandler,sear99,dicastro}
and, more recently, in colloidal science, explicitly accounting for ion
condensation around particles~\cite{kegel}.  
Numerical evidence of cluster phases has also been recently
reported~\cite{sear99,sciortino04,coniglio04,reatto04}. Interestingly, 
at low densities, particles aggregates can act as monomers of a supramolecular 
liquid, showing phenomena typical of the liquid state as crystallization and
glass transition~\cite{sciortino04}.

In this work we study an off-lattice three dimensional representative
model of particles interacting via a short range attraction (which we
model with the generalization to very large $\alpha$ of the
Lennard-Jones (LJ) $~2\alpha-\alpha~$ potential, as previously proposed by
Vliegenthart {\em et al.}~\cite{lekker}) complemented by a long range
repulsion, which we model by a screened electrostatic Yukawa
potential. Screening is included in the model, since we are mostly
interested in describing charged colloidal and protein solutions,
where addition of salt in the dispersant medium provides an efficient
mechanism of reduction of the repulsive interaction range.  Modeling
the short range attraction with the Lennard-Jones $2\alpha-\alpha$
potential is of course arbitrary, but at the same time representative
(by tuning the exponent $\alpha$) of all short-range potentials,
arising either from depletion interactions or from van der Wall
forces. For different choices of the repulsive potential parameters,
we calculate the ground state energy and structure of clusters of
different size, to evaluate the conditions under which stable cluster
phases are expected at low temperatures, where entropic effects can be
neglected. We show that, on varying the range of interaction and the
intensity of the long range repulsive potential, it is possible to
obtain stable cluster structures of different geometry.  To locate the
ground state structures and their energies, we have used a modified
version of the basin-hopping algorithm~\cite{wales97} which has been
proposed and extensively applied in models of simple liquids and
molecular systems. To extend the range of studied cluster sizes
beyond what it is possible with present numerical resources, we
complement the ground state calculations by analytic calculations of
the cluster self energy, under the assumption of spheric cluster
shape.

The paper is organized as follows. Sec.~\ref{2n-n:sec} contains a
study of the clusters ground states for the  $2\alpha-\alpha$ attractive 
potential for two values of $\alpha$, typical of short-range attractive 
interactions. A comparison of ground state energies and cluster shapes 
between the short range attractive potential and both the standard $\alpha=6$ 
case (from Ref.~\cite{wales97}) and the Hard Sphere (HS) case (from
Ref.~\cite{sloane95}) is also reported. In Sec.~\ref{competition:sec}
we present cluster energies and geometries resulting from the
addition of a repulsive Yukawa potential, for different values of the
potential parameters. Sec.~\ref{sphere:sec} is devoted to the analytic
calculation of the cluster self-energy under the assumption of
spherical shape. Finally, Sec.~\ref{conclusions:sec} contains a
discussion of the results and conclusions.
\section{Cluster Ground States for attractive short range $2\alpha-\alpha$ 
potential}
\label{2n-n:sec}
\begin{figure}[t]
\centering
\includegraphics[width=0.80\textwidth]{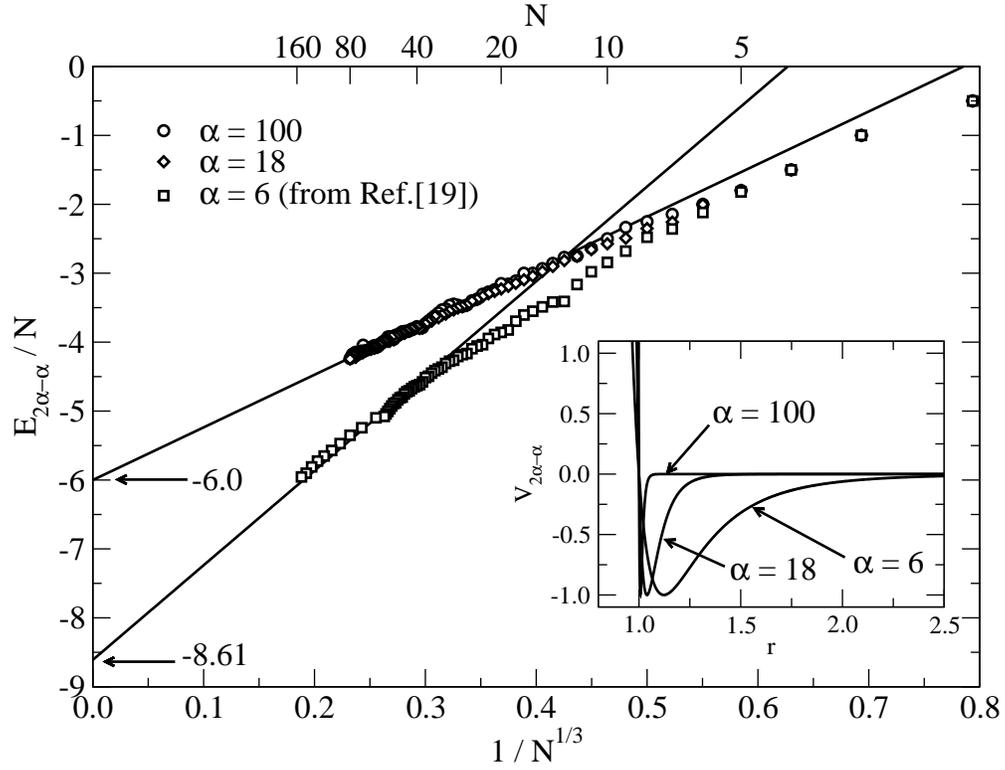}
\caption{Ground state energy per particle as a function of $1/N^{1/3}$
($N$ on the top $x$-axis) for the $2\alpha-\alpha$ potential at the
indicated values of $\alpha$. The values of the energy per particle 
for the fcc crystal structure are indicated by arrows. 
{\em Inset:} Radial dependence of interaction potential $V_{2\alpha-\alpha}$
for the three studied values of $\alpha$.}
\label{fig:potentials}
\end{figure}

Short-range attractive interactions arise in colloidal systems either
from van der Waals interactions (integrated over the volume of the
interacting particles) or via a depletion mechanism, i.e., when
particles of intermediate dimension between the suspended colloids and
the solvent molecules are added to the solution~\cite{israelovich}.
Here we model the short-range attractive potential in a numerically
convenient way with the generalization of the Lennard-Jones potential,
namely the $2\alpha-\alpha$ potential, previously proposed by
Vliegenthart {\em et al.}~\cite{lekker},
\begin{equation}
V_{2\alpha-\alpha}(r)=4 \epsilon  \bigg[ \bigg(\frac{\sigma}{r}\bigg)^{2\alpha}-
\bigg(\frac {\sigma}{r}\bigg)^{\alpha}\bigg],
\label{eq:potsr}
\end{equation}
where the value of $\alpha$, which in standard LJ case is fixed to
six, is varied to control the range of interactions.
In what follows we have chosen $\epsilon$ and $\sigma$ as the units of
energy and length respectively. We focus our attention to three values
of the parameter $\alpha$ that are relevant for our discussion.
Firstly, $\alpha=100$, which corresponds to an attractive range of the
order of a few percent of $\sigma$, typical of vary narrow-ranged
depletion interactions. Next, we consider the case $\alpha=18$, which
corresponds to the case for which the value of the potential 
energy at the distance of the second neighbor shell becomes negligible. 
Finally, for comparison, we also report results from Ref.~\cite{wales97}, 
for the Lennard-Jones case $\alpha=6$. The three potentials 
($\alpha=100, 18, 6$) are illustrated in the inset of 
Fig.~\ref{fig:potentials}.

To calculate the ground state energy of clusters of different size $N$
and their geometry, we have used a modified version of the
basin-hopping algorithm~\cite{wales97}, introduced by Wales and Doye.
Such an algorithm consists of a constant-temperature Monte-Carlo
simulation where the actual potential energy surface is transformed
into a stair-like surface by replacing the potential energy function with the
values of the potential energy of the closest local minimum,
named inherent structure~\cite{stillinger}. Acceptance criterion is thus
based upon the change in the inherent structure energy. In this way,
particles can be moved to regions with high potential energy with
large acceptance rate, facilitating the overcome of the potential
energy barriers. To further favor barrier crossing, every $100$
Monte-Carlo steps, the less bounded atom is removed and re-inserted in
the position with lowest insertion energy. The largest cluster size studied
is of the order of 80 particles.

Fig.~\ref{fig:potentials} shows the ground state energy per particle
$E/N$ as a function of $N^{-1/3}$, for all three values of $\alpha$.
Data are reported as a function of $N^{-1/3}$ since, for large
clusters, one expects to observe the functional form $E/N = c_0 +
c_1 / N^{1/3}$. Indeed, the cluster energy can be expressed as the  
sum of a bulk term, $c_0 N$, and a surface term, $c_1 N^{2/3}$.  
We find that, for $\alpha=100$, when clusters are larger than $\approx 13-15$
particles, the ground state energy per particle is well represented by
the fit $E/N = -5.937 + 7.435/N^{1/3}$, providing an estimate for the
bulk energy compatible with a close-packed fcc or hcp crystalline ordering 
value ($-6$). It is interesting to note that the $\alpha=18$ data are 
practically indistinguishable from the $\alpha=100$ case, suggesting that 
cluster ground state energies and geometries are the same for all potentials 
with $\alpha \ge 18$.

Fig.~\ref{fig:potentials} also reports the ground state energy per
particle for the standard Lennard-Jones case $\alpha=6$, from
Ref.~\cite{wales97}. As compared to the shorter range case, the ground
state energy of the LJ is lower, due to the attractive contributions
arising from second and further neighbors. The asymptotic behavior
(with $N^{-1/3}$) is approached at much larger sizes (order of $\approx 30-40$ 
particles) and suggests an extrapolation to the bulk value, consistent
with the estimate for the LJ fcc or hcp values of $-8.61$~\cite{stillinger01}.

The sign of the surface term, i.e., the sign of $c_1$, provides
information on the thermodynamic behavior of a macroscopic system at
low temperatures. Indeed, when $T$ is low, the ground state energy
becomes the relevant thermodynamic potential. The condition $c_1 > 0$
ensures that the system ground state in the thermodynamic limit is
composed by a single macroscopic bulk cluster. In the cluster-based
thermodynamic description pioneered by Hill~\cite{hill}, $c_1 > 0$
indicates that at low temperatures the system will phase separate in
a dense bulk liquid in equilibrium with a gas phase.

\begin{figure}[t]
\centering
\includegraphics[width=0.80\textwidth]{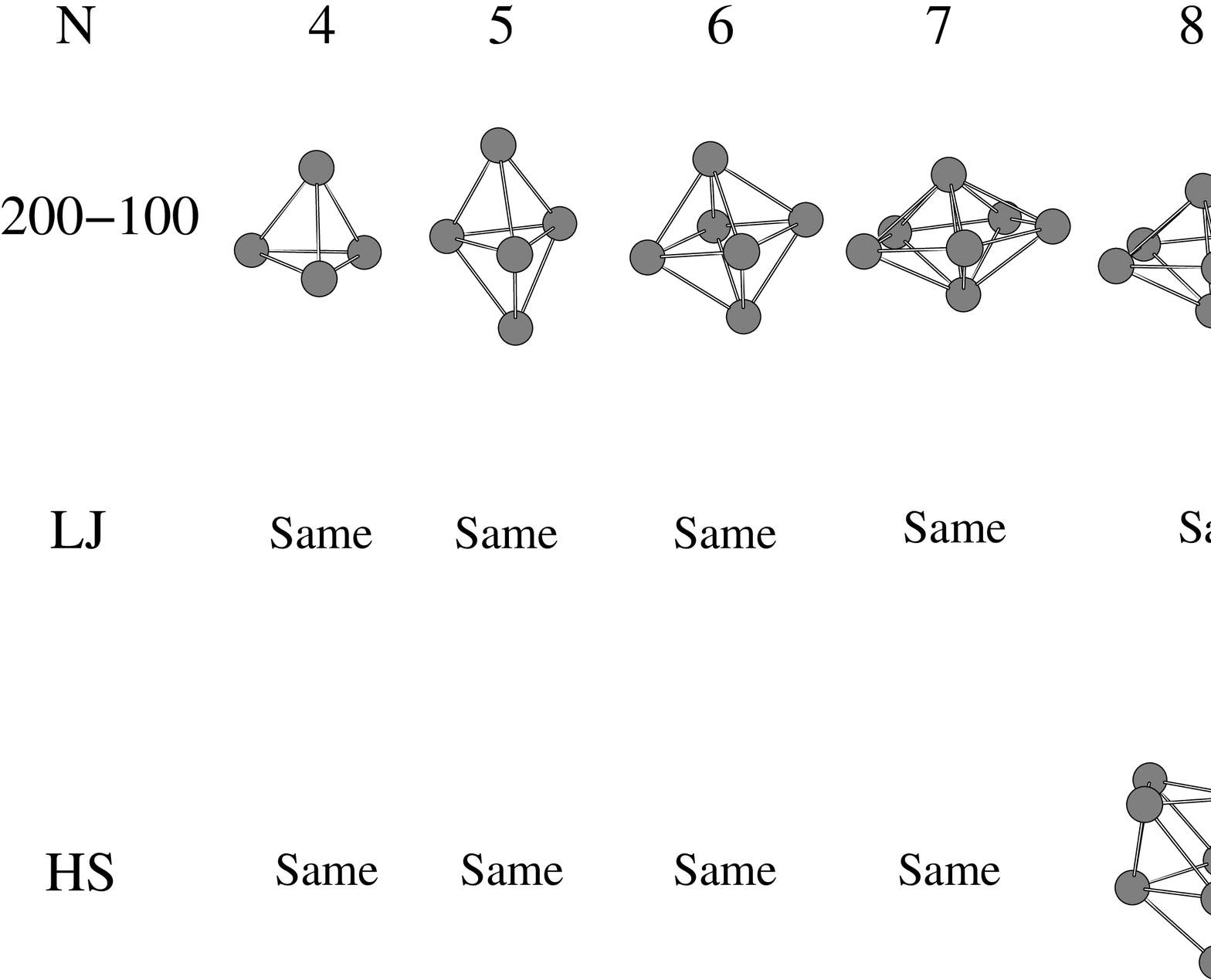}
\vspace{1.0cm}

\includegraphics[width=0.80\textwidth]{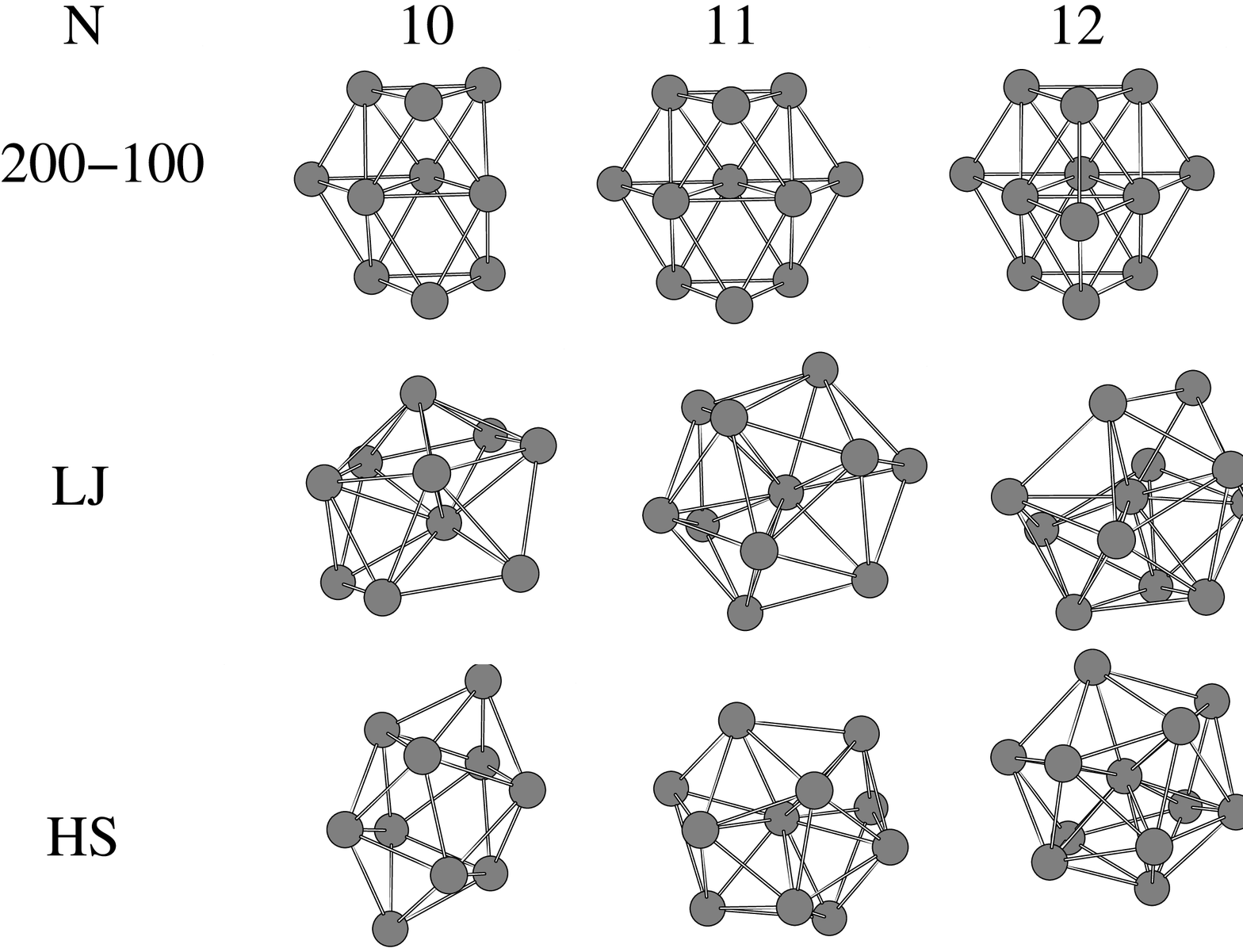}
\caption{
Cluster structures for $4\leq N \leq 13$ for the $\alpha=100, 18$ potential, 
Lennard Jones~\protect\cite{wales97}, and hard spheres~\protect\cite{sloane95}. 
`Same' indicates when the structure is identical to the $\alpha=100, 18$ case.}
\label{fig:clusters1}
\end{figure}

Next, we discuss the geometry of the ground state clusters.
Fig.~\ref{fig:clusters1} contrasts the cluster geometries for the
short-range case with the LJ case. For very small $N$, the structure 
of the clusters is independent on the range of the potential. 
For $N<10$ the structure of the clusters does not change with $\alpha$. 
For larger $N$ values, the geometry changes significantly, since the constraint 
induced by the short-range of the potential ($\alpha\ge 18$) facilitates a 
progressive layering. For example, for the case $N=13$, the icosahedron structure, 
which is particularly stable in the LJ case~\cite{wales97}, is not observed in 
the short range case.

It is instructive to compare these results also with hard spheres 
clusters geometries, calculated theoretically minimizing the second
moment of the mass distribution $M$~\cite{sloane95,lauga04}, and recently
measured with a new experimental technique introduced to produce
compact clusters of controlled numbers of small colloidal
particles~\cite{manoharan03}. For $N<11$ these clusters have
been found to be identical to those calculated in Ref.~\cite{sloane95}. 
The third row of Fig.~\ref{fig:clusters1} reports the HS clusters from 
Ref.~\cite{sloane95}. As for the comparison with the LJ case, 
for $N \le 7$ all clusters have the same structure, while for
$N >7$ differences in the potential start  to be significant.
\section{The competition between attraction and repulsion terms}
\label{competition:sec}
\begin{figure}[t]
\centering
\includegraphics[width=0.80\textwidth]{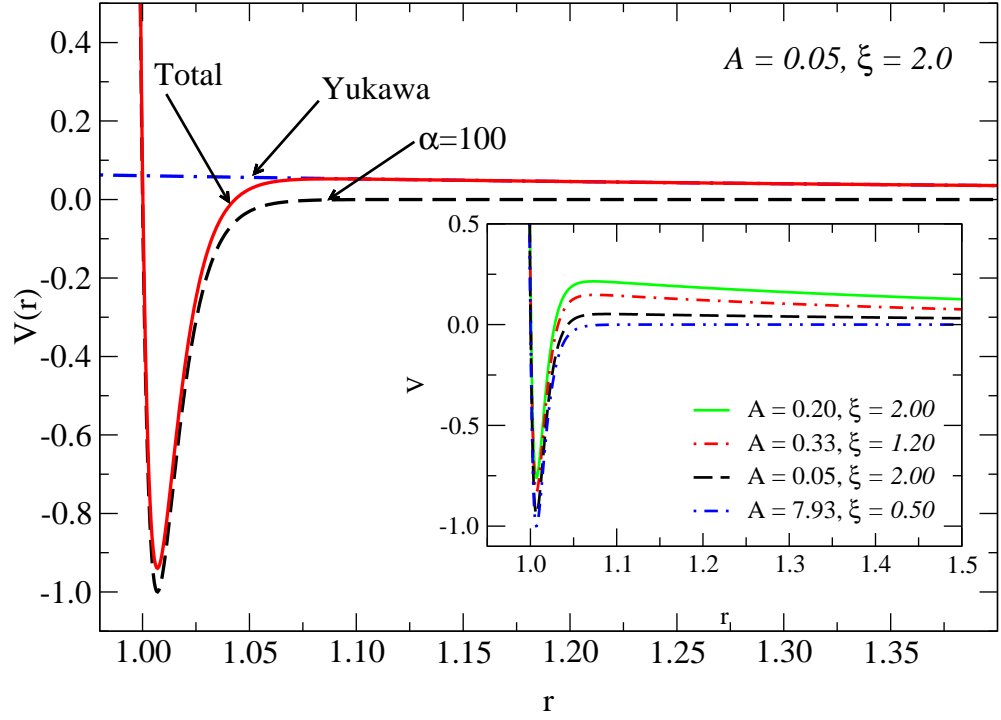}
\caption{Short range attractive $2\alpha-\alpha$ (dashed line),
long range Yukawa repulsive (dot dashed line) and total (solid line) 
potentials for $A=0.05$ and $\xi=2.0$.
{\em Inset:} $V_{2\alpha-\alpha}+V_Y$ at the indicated values of 
$A$ and $\xi$.}
\label{fig:potentials2}
\end{figure}

In this section we report calculations of the cluster ground state
energy for a potential composed by a short-range attractive part
(modeled only with $\alpha=100$, since we have shown previously that
$\alpha=18$ can not be distinguished from the $\alpha=100$ case)
complemented by a screened electrostatic repulsive interaction modeled
by a Yukawa potential,
\begin{equation}
V_{Y}(r)=A\; \frac{e^{-r/\xi}}{r/\xi}.
\label{eq:potyuk}
\end{equation}
The resulting total potential is thus $V(r)=V_{2\alpha-\alpha}+V_{Y}$. 
In the following, $A$ is given in unit of $\epsilon$ and $\xi$ in unit 
of $\sigma$. Fig.~\ref{fig:potentials2} shows the shape of the total 
potential for different $\xi$ and $A$ values.  
The resulting potential is conceptually similar to the well-known DLVO 
potential~\cite{dlvo}, even if it differs for the absence of the weak 
secondary minima and the finite value of the attractive interaction energy.

\begin{figure}[b]
\centering
\includegraphics[width=0.80\textwidth]{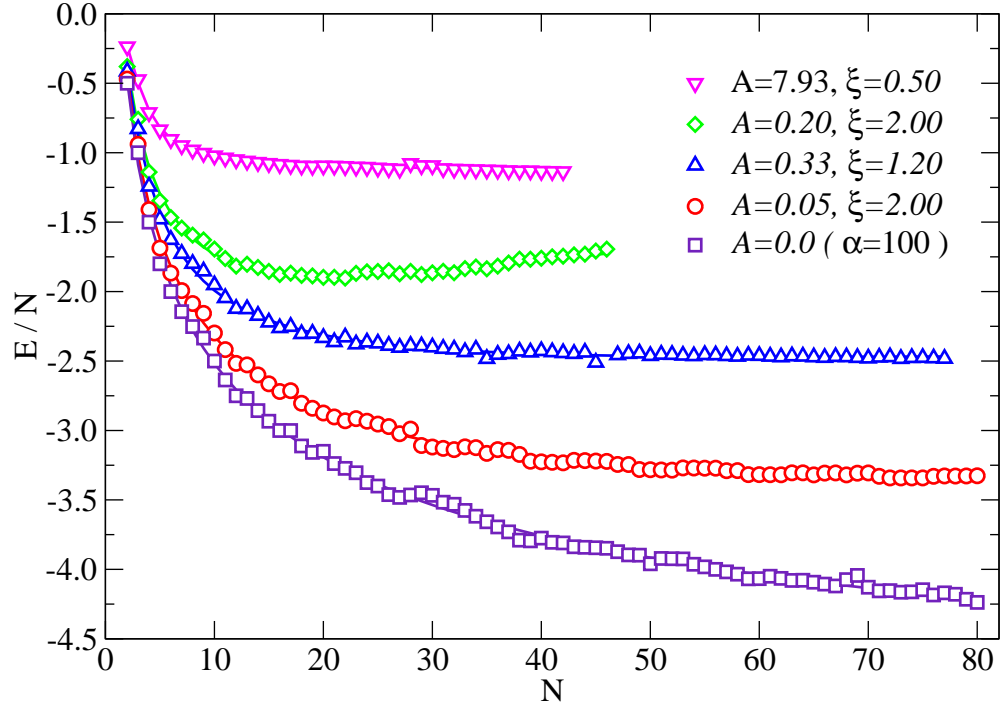}
\caption{Ground state energies per particle for five different choices of the 
potential parameters. The energies have been calculated by basin-hopping Monte 
Carlo optimization~\protect\cite{wales97}, as discussed in the text.}
\label{fig:gs-ene}
\end{figure}

Fig.~\ref{fig:gs-ene} shows the ground state energy per particle $E/N$
for different $A$ and $\xi$ values. The same minimization procedure
as described in the previous section has been implemented, with the
additional condition that, in the search procedure, moves creating
disconnected clusters are rejected.  For appropriate values of $A$ and
$\xi$ (for example $A=0.2$ and $\xi=2.0$), the cluster ground state energy 
shows a minimum at a finite size $N^*$, indicating that clusters of size
larger than $N^*$ are energetically disfavored. A simple physical
explanation of the existence of an optimal size can be given as
follows. At large enough $N$, the addition of an extra layer of
particles will contribute to the energy with a negative term due to
the attractive nearest neighbor interactions (which involves only
interactions with particles in the surface layer due to the short
range of the potential), plus a positive contribution arising from the
Yukawa repulsion which, due to its longer range, involves instead a
large fraction of particles in the cluster. The balance between these
two terms provides a condition for $N^*$. Of course, the larger the
amplitude or the longer the range of the repulsion, the smaller $N^*$
is. The fact that an optimal finite cluster size is observed,
suggests that a macroscopic system at low temperature will not form a
single aggregate, but will prefer to partition the particles into
clusters of size $N^*$. In this respect, liquid condensation is
inhibited and the structure of the system at low temperature and low
packing fractions will be constituted by a cluster phase.  At finite
temperatures, entropic contributions to the free energy will become
important and will always favor the stability of clusters of size
smaller than $N^*$. Hence, $N^*$ plays the role of an upper limit for
the cluster size. In other cases, a minimum is not found, but the
energy per particle becomes essentially flat. In these conditions, the
ground state of a macroscopic system will be composed by a highly
polydisperse cluster phase.

\begin{figure}[t]
\centering
\includegraphics[width=0.90\textwidth]{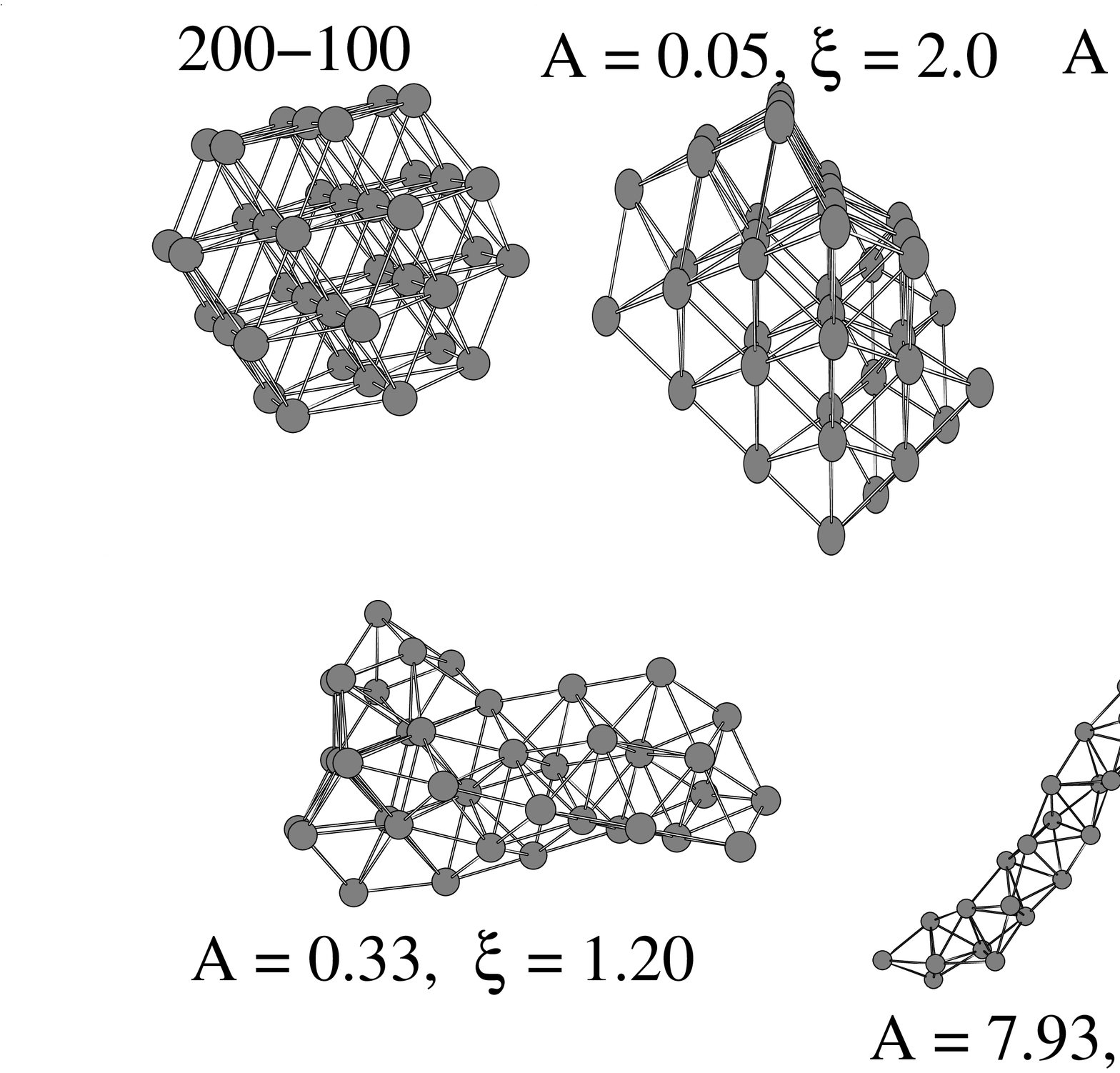}
\caption{Ground state clusters for $N=38$ for the indicated potentials.
On changing the values of the parameters it is possible to interpolate
from a close packed to an almost unidimensional ground state structure.}
\label{fig:gs-38}
\end{figure}

In Fig.~\ref{fig:gs-38} we compare the ground state cluster geometries
for $N=38$ calculated for different choices of the Yukawa parameters.  
At fixed cluster size, by changing the potential parameters the cluster
structure can be tuned from almost spherical to almost $1-$dimensional.  
The studied parameters encompass situations occurring in uncharged colloids 
($A=0$) with those typical of charged colloids ($A>0$) in apolar solvents 
or weakly screened polar solvents. 
Indeed, $A$ is proportional to the effective charge of the colloidal particle, 
while the screening length $\xi$ is controlled by the ionic strength. 
In the absence of salt, $\xi$ depends only on colloid concentration, and a small 
screening is produced by the counterions in the solvent~\cite{dlvo,likos,denton}.

\begin{figure}[t]
\centering
\includegraphics[width=0.90\textwidth]{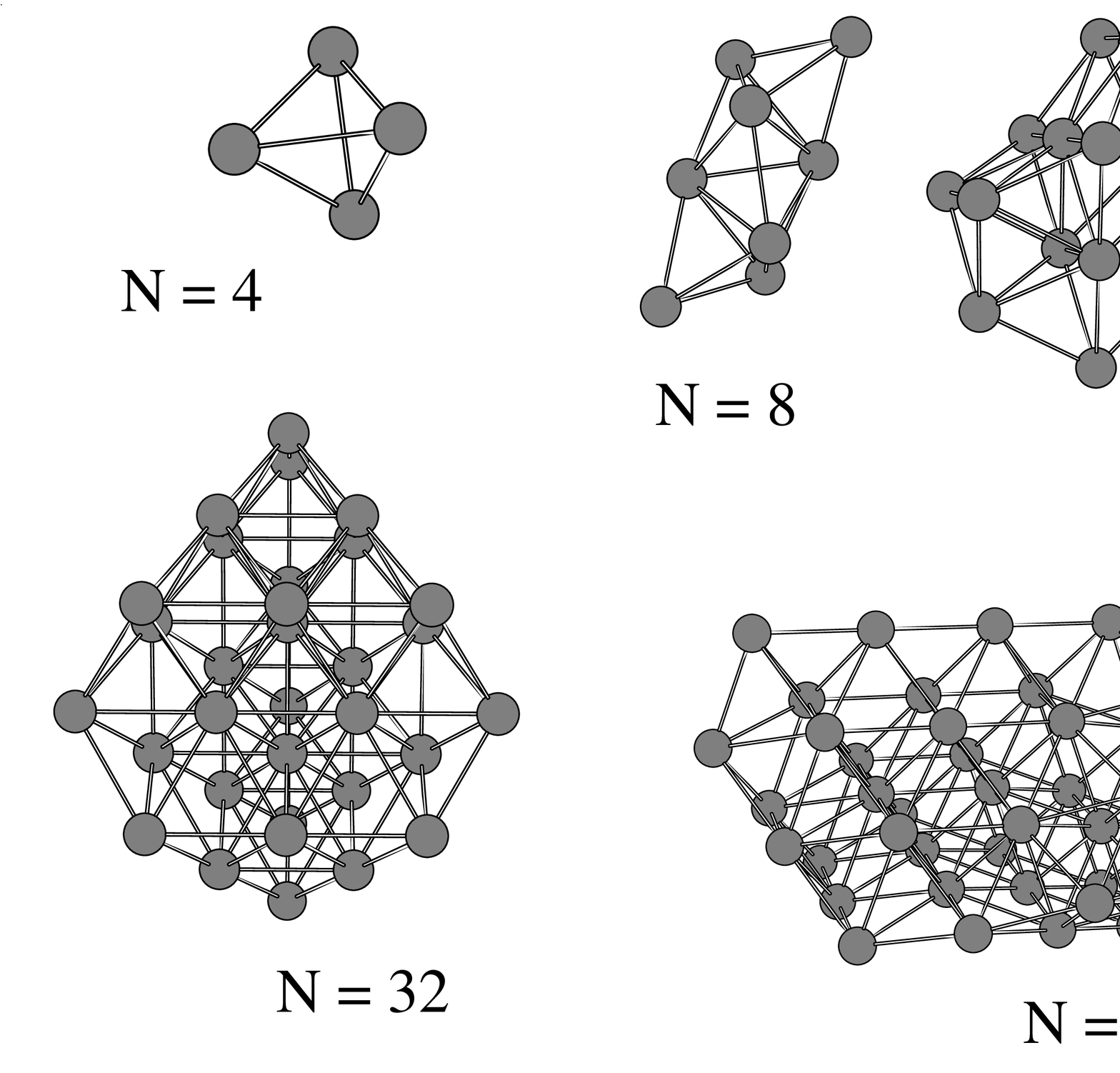}
\caption{Ground state clusters for $N=4, 8, 16, 32, 64$ for 
$A=0.05$ and $\xi=2.0$.}
\label{fig:gs-A0.05}
\end{figure}

\begin{figure}[t]
\centering
\includegraphics[width=0.80\textwidth]{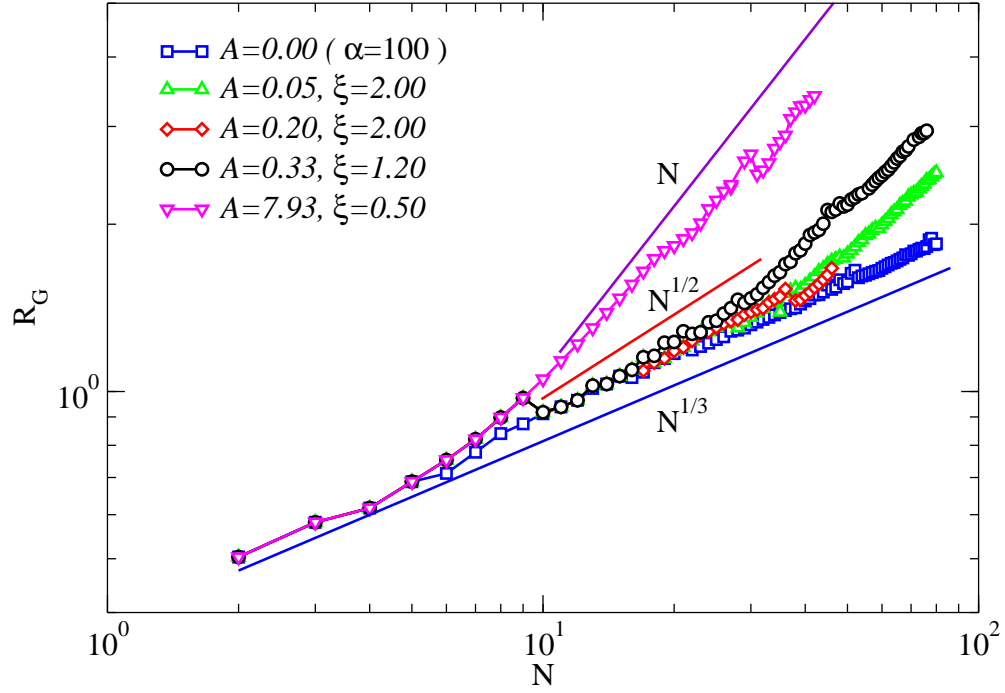}
\caption{Size dependence of the gyration radius $R_G(N)$ (in units of $\sigma$)
for all the considered potentials. The expected behaviors for spherical, planar, 
and unidimensional clusters, discussed in the text, are shown as solid lines.}
\label{fig:gyration}
\end{figure}

Fig.~\ref{fig:gs-A0.05} shows the $N$-dependence of the cluster
geometry, for the case $A=0.05$ and $\xi=2$, to highlight the changes
in the preferential geometry with cluster size. It appears that the
geometry of the clusters is strongly size dependent. To convey this
point, we show in Fig.~\ref{fig:gyration} the size dependence of the
gyration radius $R_{G}$, defined in terms of particle's coordinates
${\bf r}_i$ and center of mass coordinates ${\bf R}_{CM}$ as
\begin{equation}
R_{G} = \frac{1}{N^{1/2}} \left[\sum_{i=1}^N ({\bf r}_i 
-{\bf R}_{CM})^2\right]^{1/2}.
\label{gyration:eq}
\end{equation}
For spherical clusters, $R_{G} \sim N^{1/3}$, for planar structures,
$R_{G} \sim N^{1/2}$, and for linear structures, $R_{G} \sim N$.
Fig.~\ref{fig:gyration} shows that, while in the pure short-range
attractive case clusters retain their spherical shape for any size, in
the case of addition of a repulsive potential only clusters of small
enough size are spherical. On increasing $N$, the ground state cluster
structure becomes more and more linear. The sharp cross-over size
between spherical and linear cluster shape decreases on increasing $A$
or $\xi$.
Similarly, the linear cluster becomes thinner and thinner when repulsive
effects becomes more and more relevant, as evidentiated by the
amplitude of the linear $N$-dependence of $R_{G}$.  Once the linear
cluster shape is established, the cluster energy increases linearly
with the the cluster size, resulting in the essentially flat
$N$-dependence of the energy per particle (see Fig.~\ref{fig:gs-ene}).

Three possible scenarios appear to take place in the cluster
structure of the studied potential: {\em i)} very small $A$ 
values, where the repulsive energy is not sufficient to overcome the
attractive contribution. Under these conditions, clusters are
spherical and the standard behavior (infinite optimal size, liquid
phase) is recovered; {\em ii)} an intermediate case, where a minimum in
the cluster energy per particle is observed at an optimal size $N^*$,
with a spherical cluster structure (as, for example, the case
$A=0.2$, $\xi=2$). Linear clusters can be built but with an energy
slightly higher than that of the optimal cluster size; {\em iii)} a 
monotonically decreasing cluster energy per particle, ending into a 
flat curve, signaling equivalent stability for linear clusters of very 
different size (as, for example, the extreme case $A=7.93$, $\xi=0.5$).

It is tempting to speculate that when parameters are chosen in such a
way that linear growth is preferential, at finite but small
temperature (i.e., when structures different from the ground state
structure are also probed) clusters made of arms branching out from
regions of locally higher energy can be generated. These structures,
if macroscopic in size, would generate a gel-like structure, since no
driving force for macroscopic aggregation is present (due to its
inhibition caused by the repulsive term).
\section{The spherical case: the cross-over from infinite to finite optimal 
size} 
\label{sphere:sec}
The numerical results presented in the previous sections confirm the
possibility of creating stable clusters of optimal size, playing with
the competition between the short range attraction and the long range
repulsion. We have discussed the fact that in the limit of low
temperatures and small packing fractions, i.e., in the conditions
where entropic contributions as well as cluster-cluster interactions
can be neglected, the cluster ground state energy $E$ plays the role of
relevant thermodynamic potential and, therefore, the minimum in $E/N$
vs $N$ provides an estimate of the stable cluster size $N^*$. Under
these conditions, the system will partition into clusters of size $N^*$
and no thermodynamic (macroscopic) coagulation will take place. Hence, the
condition $N^*=\infty$ acts as critical condition for the existence of
a macroscopic phase separation as opposed to a phase made of clusters
of finite size $N^*$.

To estimate the critical line separating the liquid phase
($N^*=\infty$) from a cluster phase ($N^*$ finite) in the $(A,\xi)$
parameter space, it is necessary to calculate the ground state energy
for clusters of large size, which prevents the use of the "exact"
numerical minimization which have been presented previously.
An estimate of the critical line in the $(A,\xi)$ space can be calculated
analytically, under the hypothesis that clusters  have a spherical
shape and a homogeneous density of particles.  The assumption of a
spherical shape is expected to be valid in the region of parameter
space where $N^*=\infty$, i.e., where $A$ or $\xi$ are small and
attraction is still dominant, as shown in the previous section.  
The numerical study of the $N$-dependence of the ground state of the
$2\alpha-\alpha$ potential reported above shows that the $N$-dependence 
of the attractive part of the potential can be written as
$V_{2\alpha-\alpha}(N)/N=c_0 + c_1 N^{-1/3}$ with $c_0 \approx -6$ and 
$c_1 \approx 7.4$. Since the value of the bulk energy ($c_0$) reflects 
the energy of a very compact state (in which each particles is surrounded 
by 12 neighbors), the local density can be approximated with the local
density of the fcc structure, providing a mean to convert the cluster
radius $R$ to the cluster size using $N=\eta_{fcc} (2R/\sigma)^3$, 
with $\eta_{fcc}\simeq 0.74$.

To estimate the repulsive energy of a cluster of radius $R \gg \sigma$, we
assume for simplicity a particle-particle radial distribution function 
$g(r)=\Theta(r-\sigma)$ where $\Theta$ is the Heaviside step function. 
This choice ensures that unphysical interactions, which would arise
by pairs of points in the sphere closer than $\sigma$, are eliminated. 
A more precise calculation could be performed numerically
by implementing a more detailed expression for $g(r)$ (which, for
example, could be calculated with the Verlet-Weiss $g(r)$ for HS, or
with the known $g(r)$ of the fcc structure). The simpler step-wise
approximation used here is sufficient to predict the shape of the
critical line. For large cluster sizes, the resulting expression for 
the cluster repulsive energy per particle is 
(see Appendix~\ref{appendixcal})
\begin{eqnarray}
\frac{E_Y(R)}{N} =
\frac{3 A \eta_{fcc}}{2 \sigma^3} \xi^2\,
    \left[ \frac{-24\,\xi^2\,
         {\left( R + \xi \right) }^2}{e^
          {\frac{2\,R}\xi}\,R^3} + 
      \frac{16\,\left( \sigma + \xi \right) }
       {e^{\frac{\sigma}\xi}} - 
      \frac{12\,\left( {\sigma}^2 + 2\,\sigma\,\xi + 
           2\,\xi^2 \right) }{e^
          {\frac{\sigma}\xi}\,R} \right. \nonumber \\
      + \left.\frac{{\sigma}^4 + 4\,{\sigma}^3\,\xi + 
         12\,{\sigma}^2\,\xi^2 + 
         24\,\sigma\,\xi^3 + 
         24\,\xi^4}{e^
          {\frac{\sigma}\xi}\,R^3} \right].
\label{WR-yuk:eq}
\end{eqnarray}

\begin{figure}[t]
\centering
\includegraphics[width=0.80\textwidth]{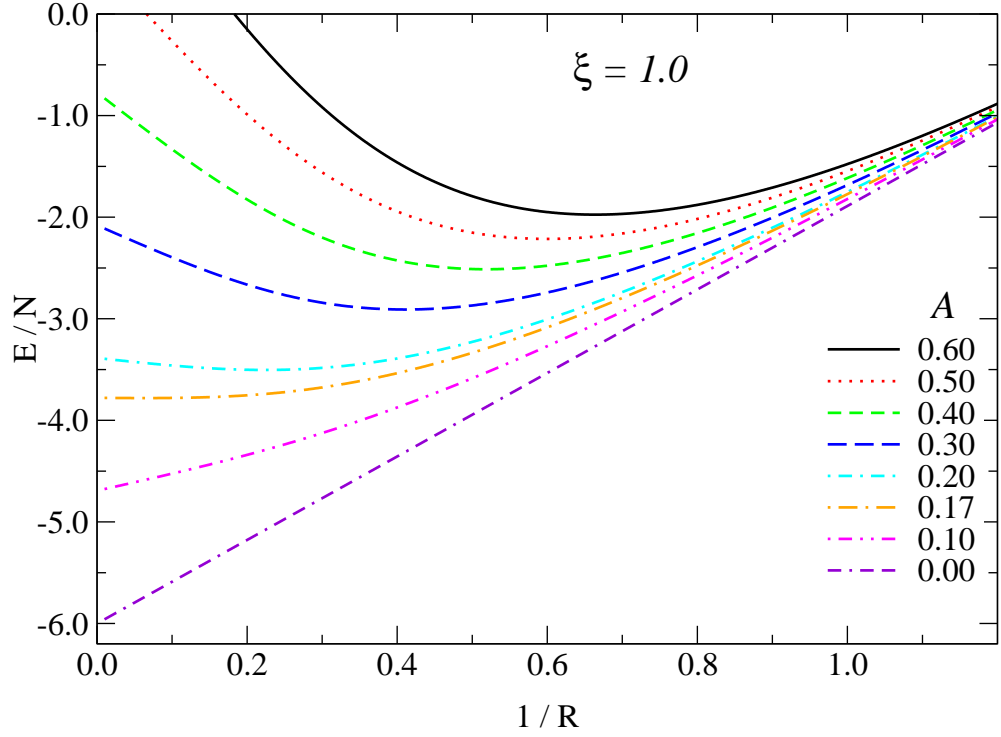}
\caption{Energy per particle in the spherical approximation for $\xi=1.0$, at the 
indicated values of $A$. A minimum starts to develop for values of $A$ larger 
than a critical value $A_c(\xi)$.
}
\label{fig:sphere1}
\end{figure}

\begin{figure}[t]
\centering
\includegraphics[width=0.80\textwidth]{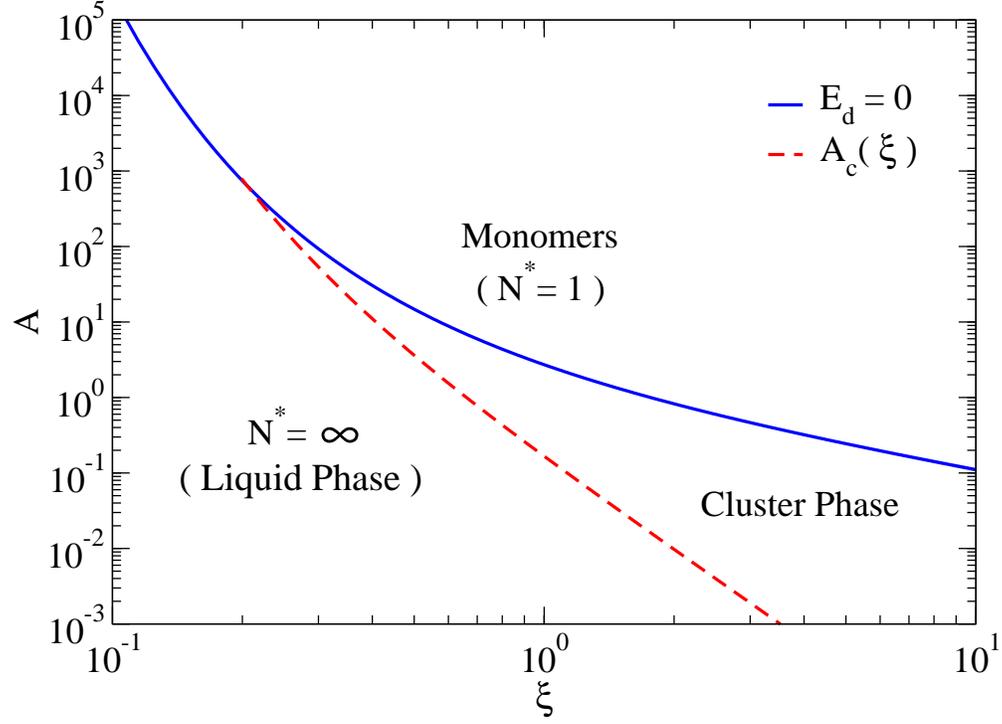}
\caption{Critical line $A_c(\xi)$ separating the $N^*=\infty$ (liquid) 
from the cluster phase on the $(A,\xi)$ parameter plane, together with 
the line of zero dimer energy $E_d$, separating clusters 
(either of finite or infinite size) from a phase of stable monomers.}
\label{fig:phasediagram}
\end{figure}

Fig.~\ref{fig:sphere1} shows the $1/R$ dependence of the cluster
energy per particle, i.e., the sum of the attractive and the repulsive
part, for $\xi=1$ and several values of $A$. On increasing $A$, a
progressive bending up of the curves takes place for large cluster
radii, until for a critical value $A_c$ ($\approx 0.17$ for $\xi=1$),  
the slope of  $E(R)$ becomes flat at $R = \infty$, signaling that the lowest 
ground state does not require that  all particles belong to the same cluster. 
For $A>A_c$, a minimum of $E(R)$ arises at  finite $R$ value. 
To estimate the critical value of $A(\xi)$, the expression Eq.~(\ref{WR-yuk:eq})
for $E_Y/N$ can be expanded in powers of $1/R$. 
At first order in $1/R$ we find, 
\begin{equation}
\frac{E_Y(R)}{N} \approx 
24\,A\,\eta_{fcc}  \frac{{\xi}^3}{\sigma^3}
     \left( \frac{\sigma}{\xi} + 1 \right) e^
     {-\frac{\sigma}{\xi}} - 
      18 \,A\,\eta_{fcc}  \frac{{\xi}^4}{\sigma^3}
     \left( \frac{{\sigma}^2}{\xi^2} + 2 \frac{\sigma}{\xi} + 
       2 \right) e^
      {-\frac{\sigma}{\xi}}\frac{1}{R},
\end{equation}  
which provides, by adding the attractive part (also linear in $1/R$) and setting  
the resulting coefficient of the $1/R$ part to zero, $A_c$ as a function of $\xi$,
\begin{equation}
A_c(\xi)=\frac{c_1}{36 \eta^{4/3}} \frac{\sigma^4}{\xi^4} 
\frac{\xi^2}{2\xi^2 + 2\xi \sigma+\sigma^2} 
e^{\frac{\sigma}{\xi}}.
\label{axicritic:eq}
\end{equation}
The critical line $A_c(\xi)$, shown in Fig.~\ref{fig:phasediagram} in the $(A,\xi)$ 
plane, locates the region where, at low temperature, a phase of clusters of finite 
size is expected. For $\xi > \sigma$,  $A_c \approx \xi^{-4}$.   
 
Finally, to provide an estimate of the region where, even at $T=0$ monomers 
are the stable state, we show in Fig.~\ref{fig:phasediagram} the line 
corresponding to the condition of the energy of a dimer $E_d$ being zero, 
i.e., along this line the pair potential repulsive energy at distance $\sigma$ 
is equal to the  short-range attractive energy 
(i.e., $A e^{-\sigma/\xi}/{(\sigma/\xi)}=1$). 
Crossing the $E_d=0$ line from below, the dimer ground state energy 
goes from negative to positive values. The intersection of this line 
with the $A_c(\xi)$ line suggests that, for $\xi \lesssim 0.2$, 
there is no possibility of a cluster phase, with clusters larger than monomers. 
Therefore, for such small values of $\xi$, the ground state is given by an 
infinite cluster when the dimer energy is negative, crossing sharply to a solution 
of monomers when the dimer energy becomes slightly repulsive. 
Physical realization of cluster phases should then be searched in systems where the 
screening length is comparable to the particle size, i.e., in weakly polar solvents 
or for small colloidal particles.  

In concluding this section, it is worth recalling that, as seen in the
previous section, for sizes comparable or larger than $N^*$, the shape
of the cluster becomes essentially linear and the spherical cluster
calculations lose progressively their meaning. Hence, the spherical
calculations reported here should be limited to the case $N^*=\infty$, 
i.e., below the critical $A_c(\xi)$ line. Close to the critical line 
and above, more refined calculations accounting for almost linear 
cluster shapes should be performed.
\section{Conclusions} 
\label{conclusions:sec}
In this paper, we have studied, for a model where short-range
attraction and long-range repulsion are simultaneously present, 
the structure and the energy of the ground state for clusters of
different sizes. In these systems, microphase separated states
--- in the form of spherical (micellar) or columnar phases --- 
arise spontaneously, due to the introduction of a new characteristic 
length provided by the balance between attractive and repulsive
energies. Under the hypothesis of spherical clusters, an analytic
evaluation of the cluster energy has been performed, providing a
criterium for the existence of a cluster phase. 

By varying the two parameters of the repulsive Yukawa potential,
controlling respectively the amplitude of the repulsion and the
screening length, clusters of different morphology, from almost
spherical --- for small $A$ and $\xi$ values --- to almost one-dimensional
--- for larger $A$ and $\xi$ values --- can be generated. 
The evaluation of the ground state energy as a function of the cluster size 
gives evidence of a progressive tendency toward one-dimensional growth 
for all cases leading to cluster phases. 
This preferential one-dimensional growth is expected to enhance the stability 
of collective ordering into columnar or lamellar phases when cluster-cluster 
interactions are taken into account, both energetically 
and entropically~\cite{onsager}, in full agreement with the predictions for 
unscreened repulsive
interactions~\cite{groussonpre,lowprl,schmalianprl,chandler,sear99,dicastro}.

It is tempting to connect the one-dimensional growth followed by a
dynamical arrest phenomena, which is observed in several protein
solution systems~\cite{taco1,taco2,renard,emoglobinas}, to the results
discussed in the present article. Indeed, for these protein solutions,
a change in temperature or in the solvent properties can trigger an
aggregation process of proteins into cylindric clusters. Under
appropriate concentration conditions, these one-dimensional clusters
further associate to form a macroscopic gel. The mechanisms
discussed in this manuscript account for both the formation of
cylindrical clusters and the insurgence of an effective vanishing
surface tension, a condition necessary to stabilize a gel-phase
with respect to phase separation.

Finally, we want to stress that in this study we have focused on
isolated cluster properties, and attempted to connect the cluster
properties to the formation of a cluster phase, as opposed to a
condensation of a dense liquid. In doing so, we have neglected the
cluster-cluster interactions, which will play a very relevant role
also at low temperatures, for packing fractions at which the
cluster-cluster distance becomes comparable with the screening length.
This will bring into play not only thermodynamic considerations but
also, due to the low temperature, kinetic considerations. Preliminary
work in this direction indeed suggests that slow-dynamics phenomena,
related to the cluster-cluster repulsive interactions, may play a
relevant role in arresting the equilibration of systems of particles
interacting with the type of potential studied in this work~\cite{sciortino04}.
\begin{acknowledgements}
Authors acknowledge support from MIUR Cofin 2002, FIRB and MRTN-CT-2003-504712. 
We thank A. Scala and J. Groenewold for helpful discussions. 
\end{acknowledgements}
\newpage
\appendix*
\section{Calculation of the Yukawa cluster energy}
\label{appendixcal}
To evaluate the cluster energy, under the hypothesis of a homogeneous spherical 
cluster of number density $\rho$, we proceed in two steps: 
First we evaluate the potential energy of a particle located at 
distance $x$ from the center of the sphere of radius $R$, by integrating all 
contributions from points located at the same distance $r$ from the selected particle. 
Second, we evaluate the cluster energy by summing over all particles located 
at distance $x$, integrating $x$ from $0$ to $R$.
The energy of a particle located in $x$ inside a sphere of radius $R$, $W(R,x)$, 
can be written as 
\begin{equation}
W(R,x)=\rho\left[ 4 \pi \int_0^{R-x}  dr r^2 V_Y(r) g(r) + 
\int_{R-x}^{R+x} dr S(r;R,x) V_Y(r) g(r)\right],
\end{equation}
where $S(r;R,x)$ is the surface generated by the intersection of two spheres of radius $r$ 
and $R$, whose centers are located at distance $x$ apart. From standard geometry,
\begin{equation}
S(r;R,x)=2 \pi r \left[ \frac{x^2 - r^2 + R^2}{2x} - (x-r)\right].
\end{equation}
The condition $g(r)=\Theta(r-\sigma)$ acts in the integration limits 
providing two solutions, one for the case $R-x<\sigma$ and one for the
case $R-x> \sigma$, 
\begin{equation} 
W(R,x)=\left\{
\begin{array}{ll}
\rho \int_{\sigma}^{R+x}dr S(r;R,x) V_Y(r) & x > R-\sigma\\
&\\ 
\rho \left[4 \pi \int_\sigma^{R-x} dr r^2 V_Y(r) + 
\int_{R-x}^{R+x} dr S(r;R,x) V_Y(r)\right] & x < R-\sigma
\end{array}
\right..
\end{equation}
The total potential energy of the cluster $E_Y(R)$ is thus calculated as
\begin{equation}
E_Y(R)=4 \pi \rho \int_0^R dx x^2 W(R,x). 
\end{equation}
For the case of the Yukawa potential, standard integration provides the
result for $E_Y(R)$ reported in Eq.~(\ref{WR-yuk:eq}). Indeed, after
converting to energy per particle and using $\frac{\pi}{6} \sigma^3 \rho 
= \eta_{fcc}$, one finds,
\begin{eqnarray}
\frac{E_Y(R)}{N} =
\frac{3 A \eta_{fcc}}{2 \sigma^3} \xi^2\,
    \left[ \frac{-24\,\xi^2\,
         {\left( R + \xi \right) }^2}{e^
          {\frac{2\,R}\xi}\,R^3} + 
      \frac{16\,\left( \sigma + \xi \right) }
       {e^{\frac{\sigma}\xi}} - 
      \frac{12\,\left( {\sigma}^2 + 2\,\sigma\,\xi + 
           2\,\xi^2 \right) }{e^
          {\frac{\sigma}\xi}\,R} \right. \nonumber \\
      + \left.\frac{{\sigma}^4 + 4\,{\sigma}^3\,\xi + 
         12\,{\sigma}^2\,\xi^2 + 
         24\,\sigma\,\xi^3 + 
         24\,\xi^4}{e^
          {\frac{\sigma}\xi}\,R^3} \right]
\end{eqnarray}
\newpage
\end{document}